\documentclass[preprintnumbers,superscriptaddress,showkeys,byrevtex]{revtex4}
\usepackage{amsmath,amsfonts,amssymb,amscd,amsxtra,amsthm}
\usepackage{graphicx}
\usepackage{bm}
\usepackage{epstopdf}
\usepackage{multirow}
\begin{document}
\preprint{INHA-NTG-04/2012}
\preprint{KIAS-P12053}
\title{Contribution of $N^*$ and $\Delta^*$ resonances in $K^*\Sigma(1190)$ photoproduction }
\author{Sang-Ho Kim}
\email[E-mail: ]{sanghokim@ihha.edu}
\affiliation{Department of Physics, Inha University, Incheon 402-751, Republic of Korea}  
\author{Seung-il Nam}
\email[E-mail: ]{sinam@kias.re.kr}
\affiliation{School of Physics, Korea Institute for Advanced Study (KIAS),
  Seoul 130-722, Republic of Korea} 
\author{Atsushi Hosaka}
\email[E-mail: ]{hosaka@rcnp.osaka-u.ac.jp}
\affiliation{Research Center for Nuclear Physics (RCNP), Osaka 567-0047, Japan}
\author{Hyun-Chul Kim}
\email[E-mail: ]{hchkim@inha.ac.kr}
\affiliation{Department of Physics, Inha University, Incheon 402-751,  Republic of Korea}
\date{\today}
\begin{abstract}
 In this talk, we report theoretical studies on the 
 $K^{*0}\Sigma^+(1190)$ photoproduction in the tree-level Born
 approximation, employing the effective Lagrangian method. We present 
 the energy and angular dependences of the cross sections. It turns
 out that the $N^*$ and $\Delta^*$ resonance contributions are 
 negligible in the vicinity of the threshold. On the contrary, we
 observe that the $\kappa$ and $K$ exchanges in the $t$ channel and
 $\Delta(1232)$ in the $s$ channel dominate the scattering process,
 reproducing the experimental data qualitatively well. 
\end{abstract}
\maketitle
\section{Introduction}
\begin{figure}[t]
\begin{center}
\includegraphics[width=15cm]{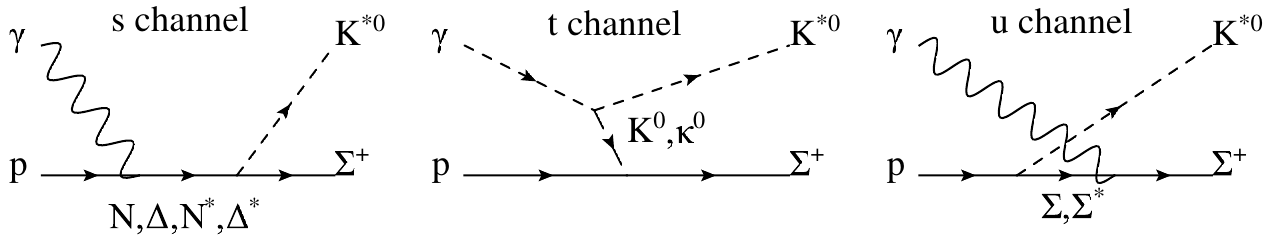}
\caption{Relevant Feynman diagrams for $\gamma p \to
  K^{*0}\Sigma^+(1190)$.}     
\label{FIG1}
\end{center}
\end{figure}

The strangeness production from various scattering processes plays an
important role in understanding the microscopic mechanism of the
strong interaction beyond the chiral limit $(m_u\sim m_d\ll m_s)$ and
extends our knowledge into multi-strangeness particles. In the present
talk, we would like to report theoretical studies on the $K^*$
photoproduction off the proton target, i.e. $\gamma p \to
K^{*0}\Sigma^+(1190)$. There have been two experimental data from the
CLAS collaboration at Jefferson laboratory~\cite{Hleiqawi:2007ad} and
the TAPS collaboration at CBELSA~\cite{Nanova:2008kr}. Theoretical
studies are also carried out based on the effective Lagrangian 
method~\cite{Oh:2006hm} and chiral quark model~\cite{Zhao:2001jw}. In
the present work for the $\gamma p \to K^{*0}\Sigma^+(1190)$ process,
we employ the effective Lagrangian method at the tree-level Born
approximation, including the $N^*$ and $\Delta^*$ resonance
contributions, such as $F_{15}(2000)$, $D_{13}(2080)$, $G_{17}(2190)$,
$D_{15}(2200)$, $F_{35}(2000)$, $G_{37}(2200)$, and
$F_{37}(2390)$~\cite{Nakamura:2010zzi}. The relevant Feynman diagrams
preserving the gauge invariance
are given in Fig.~\ref{FIG1}, in which $N$, $\Delta$, 
$\Sigma$, and $\Sigma^*$ indicate the nucleon, $\Delta(1232)$,
$\Sigma(1190)$, and $\Sigma^*(1385)$, respectively.
The strong coupling constants for the ground-state baryons $B$, 
i.e. $g_{K^*\Sigma B}$, are determined by experimental
information~\cite{Nakamura:2010zzi} as well as the Nijmegen soft-core
potential model (NSC97a)~\cite{Stoks:1999bz}. As for the strong
couplings for the baryon resonances $B^*$, we make use of the
following relation: 
\begin{equation}
\label{eq:GAMMA}
\Gamma(B^*\to K^*\Sigma)=\sum_{l,s}|G(l,s)|^2,
\end{equation}
where the amplitude $G(l,s)$ is computed by the SU(6) quark
model~\cite{Capstick:1998uh}. For simplicity, we consider only the
low-lying resonance states, since we are interested in the vicinity of
the threshold. The magnetic transition strengths for the resonances,
$\gamma N\to B^*$, are estimated from the experimental and theoretical
values for the helicity amplitudes $A(1/2)$ and
$A(3/2)$~\cite{Nakamura:2010zzi,Capstick:1992uc}. The scattering
amplitude can be written with the phenomenological form factors that
satisfy the Ward-Takahashi identity as follows: 
\begin{eqnarray}
\label{eq:AMP}
\mathcal{M}=
[\mathcal{M}_{s(p)}+\mathcal{M}_{u(\Sigma)}]F_c^2 
+ \mathcal{M}_{t(K)}F_K^2 + \mathcal{M}_{t(\kappa)} F_\kappa^2 
+ \mathcal{M}_{s(\Delta)} F_\Delta^2
+\mathcal{M}_{u(\Sigma^*)} F_{\Sigma^*}^2
+ \mathcal{M}_{s(N^*)}F_{N^*}^2
+ \mathcal{M}_{s(\Delta^*)}F_{\Delta^*}^2   
\end{eqnarray}
and the form factors are defined generically as
\begin{equation}
\label{eq:FF}
F_c=F_pF_{\Sigma}-F_p-F_{\Sigma}, 
\qquad
F_\Phi=\frac{\Lambda^2_{\Phi}-M^2_\Phi}{\Lambda^2_\Phi-q^2},
\qquad
F_B=\frac{\Lambda^4_B}{\Lambda^4_B+(q^2-M^2_B)^2},
\end{equation}
where $q$, $\Lambda_\Phi$ and $\Lambda_B$ stand for the momentum
transfer, the cutoff masses for the meson-exchange and baryon-pole
diagrams, respectively. For the details of the theoretical framework, 
readers can refer to~\cite{Oh:2006hm,Kim:2011rm}.  

\section{Numerical results}
\begin{figure}[h]
\begin{tabular}{cc}
\includegraphics[width=8.5cm]{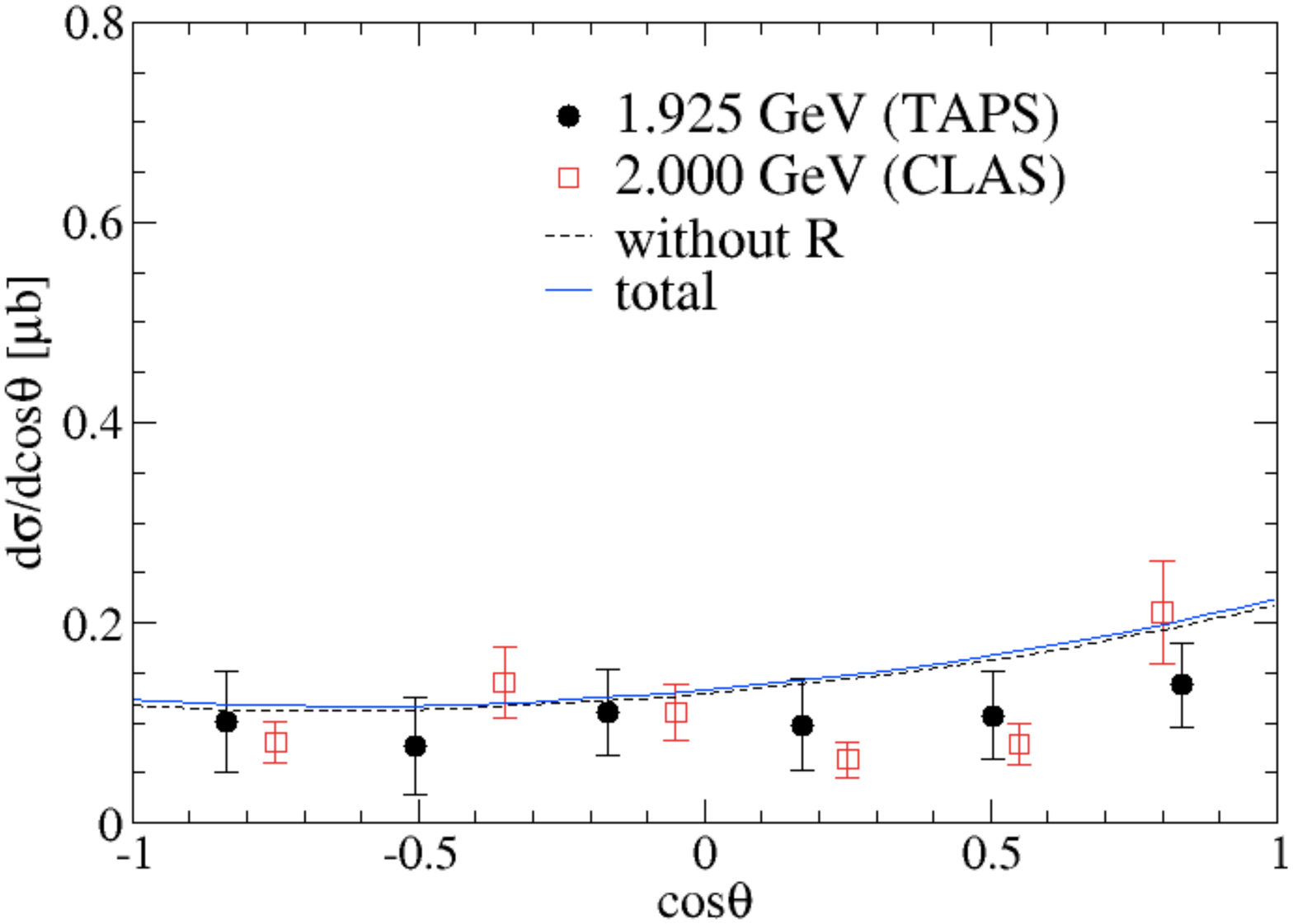}
\includegraphics[width=8.5cm]{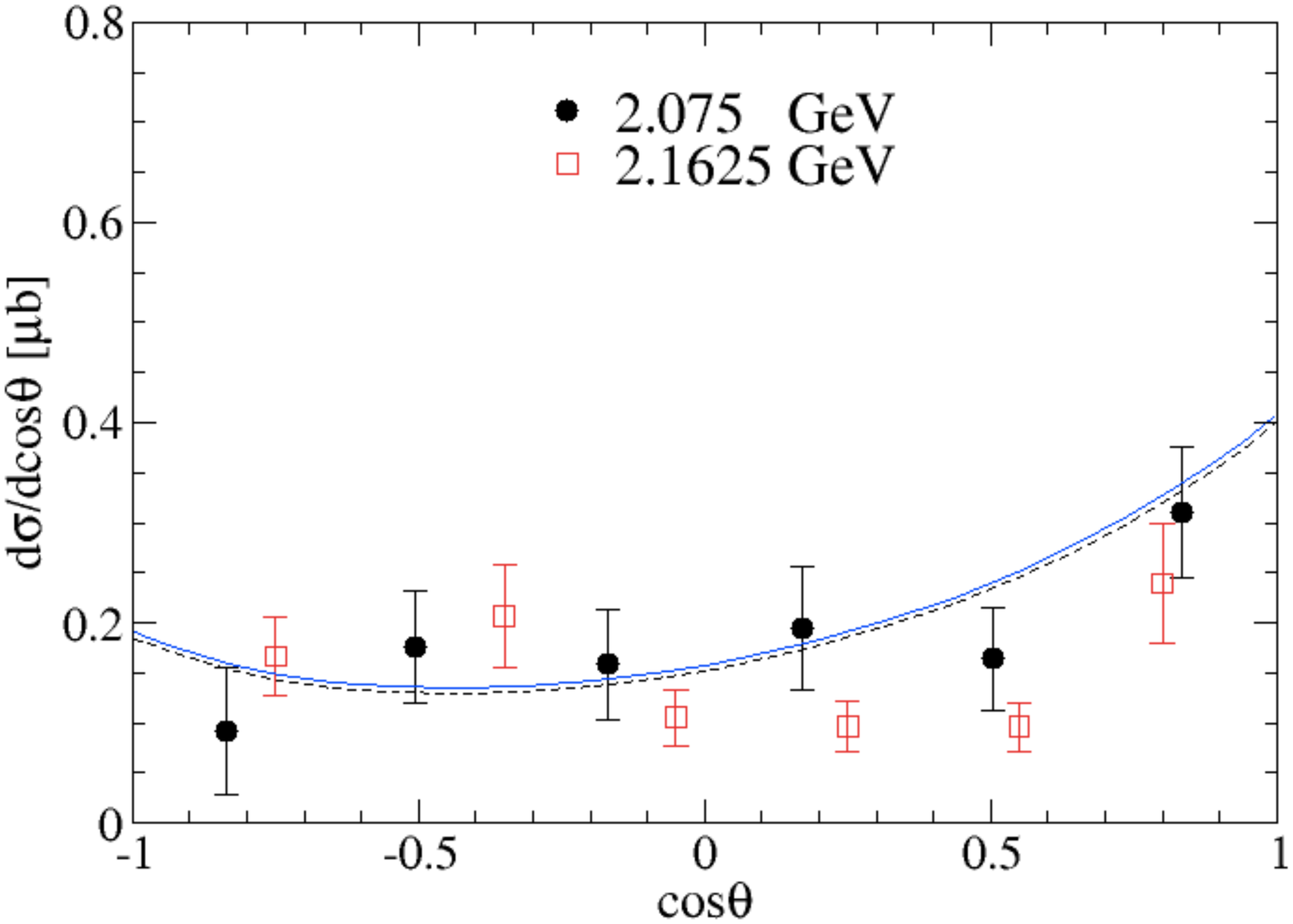}
\vspace{1.2em}
\end{tabular}
\begin{tabular}{cc}
\includegraphics[width=8.5cm]{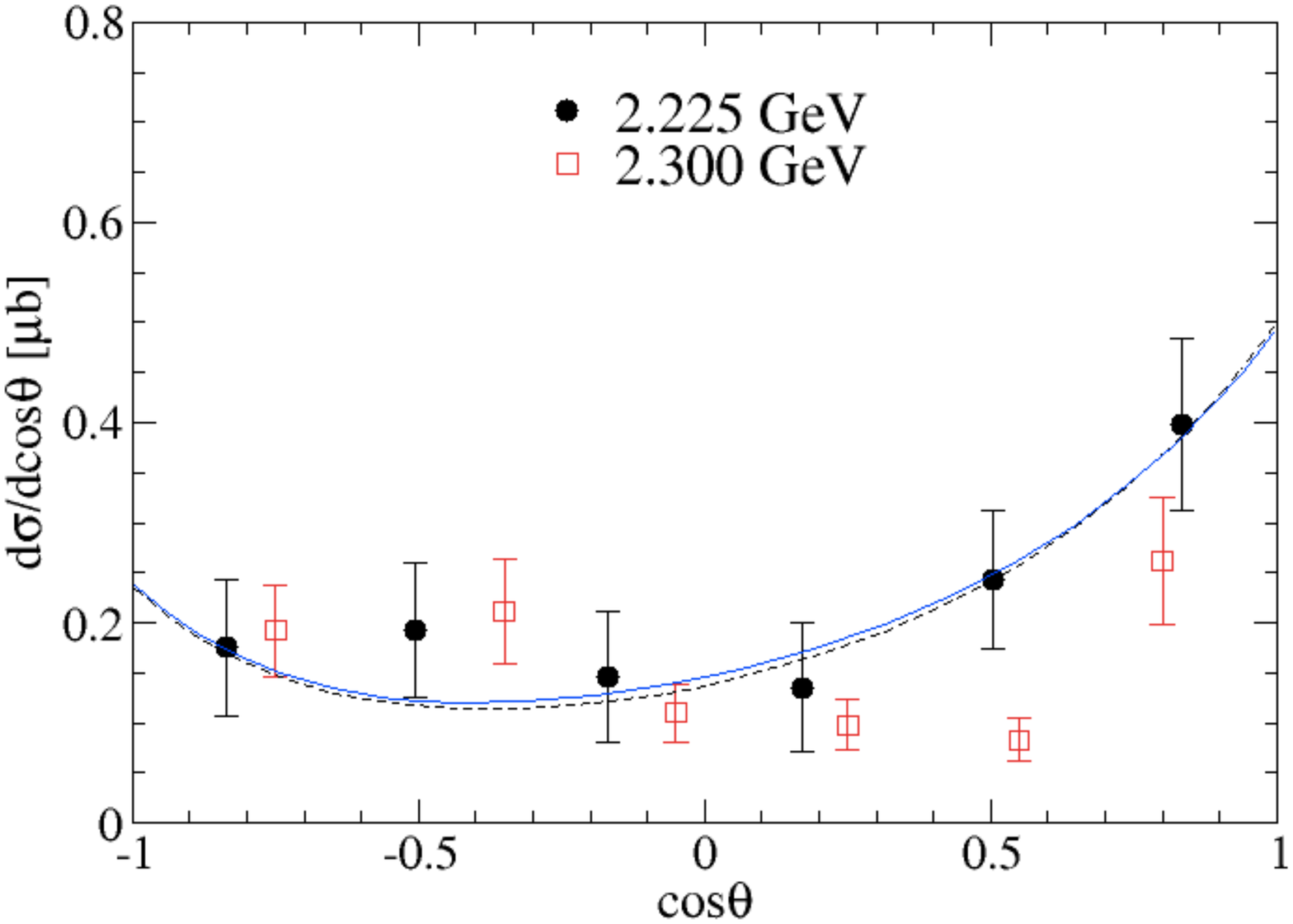}
\includegraphics[width=8.5cm]{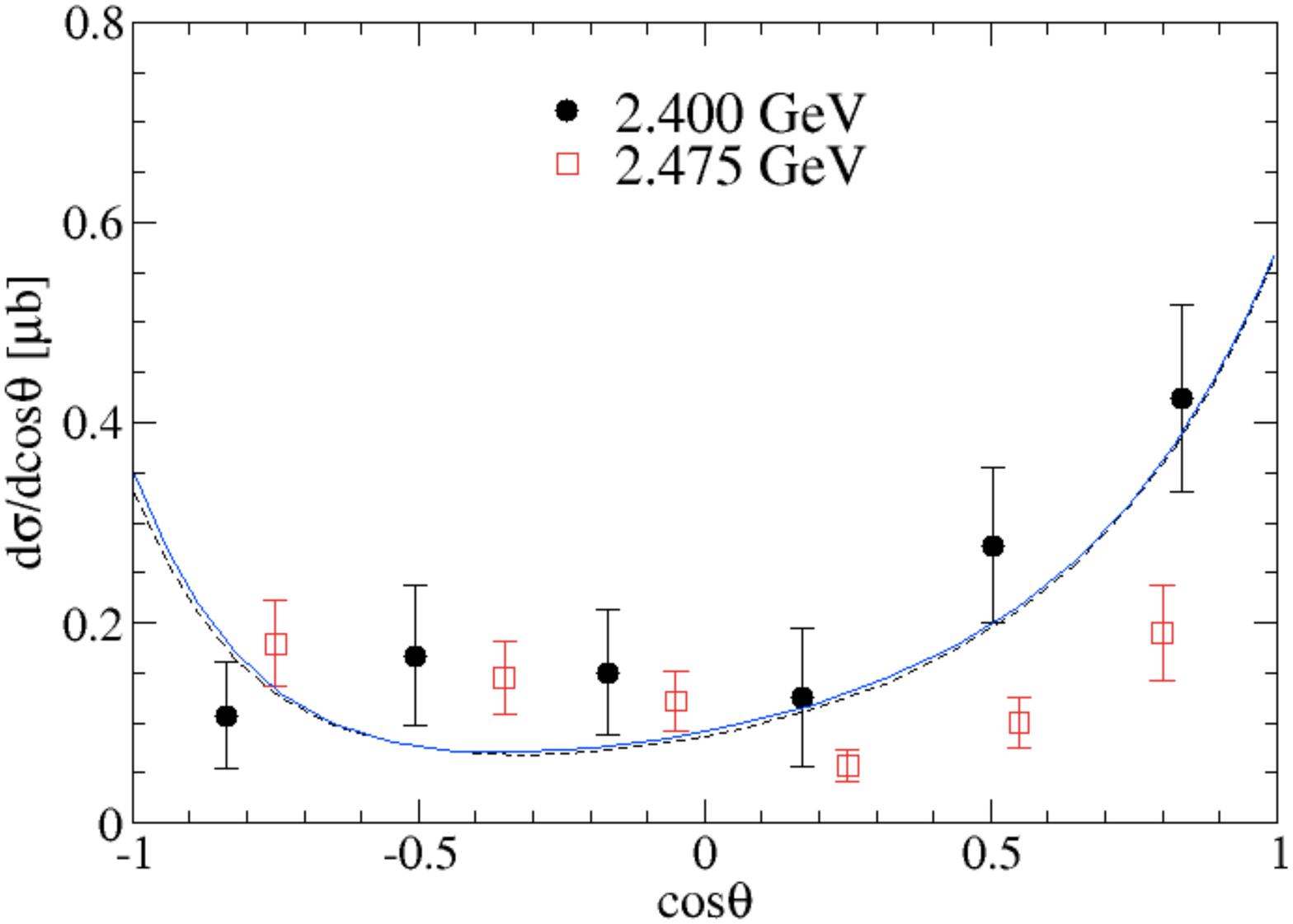}
\end{tabular}
\caption{(Color online) Differential cross section as functions of 
  $\cos\theta$ for different photon energies
  $E_\gamma=(1.925\sim2.475)$ GeV. The experimental data are taken 
  from the TAPS~\cite{Nanova:2008kr} and CLAS~\cite{Hleiqawi:2007ad}.}        
\label{FIG2}
\end{figure}

In this section, we discuss the numerical results. In Fig.~\ref{FIG2},
we show them for the differential cross section as functions of
$\cos\theta_{K^*}$ for different photon energies
$E_\gamma=(1.925\sim2.475)$ GeV. The solid 
and dotted curves indicate the cases with and without the resonance
contributions, respectively. The experimental data are taken from the
CLAS~\cite{Hleiqawi:2007ad} and TAPS~\cite{Nanova:2008kr}. As shown in
Fig.~\ref{FIG2}, the two experimental data are qualitatively well
reproduced and the resonance contributions are almost negligible. We
verify that the $K$ and $\kappa$ exchanges provide strong contribution
in the forward scattering region, whereas the backward scattering
regions are dominated by the $\Sigma(1190)$ exchange in the $u$
channel.   

\begin{figure}[h]
\begin{tabular}{cc}
\includegraphics[width=8.5cm]{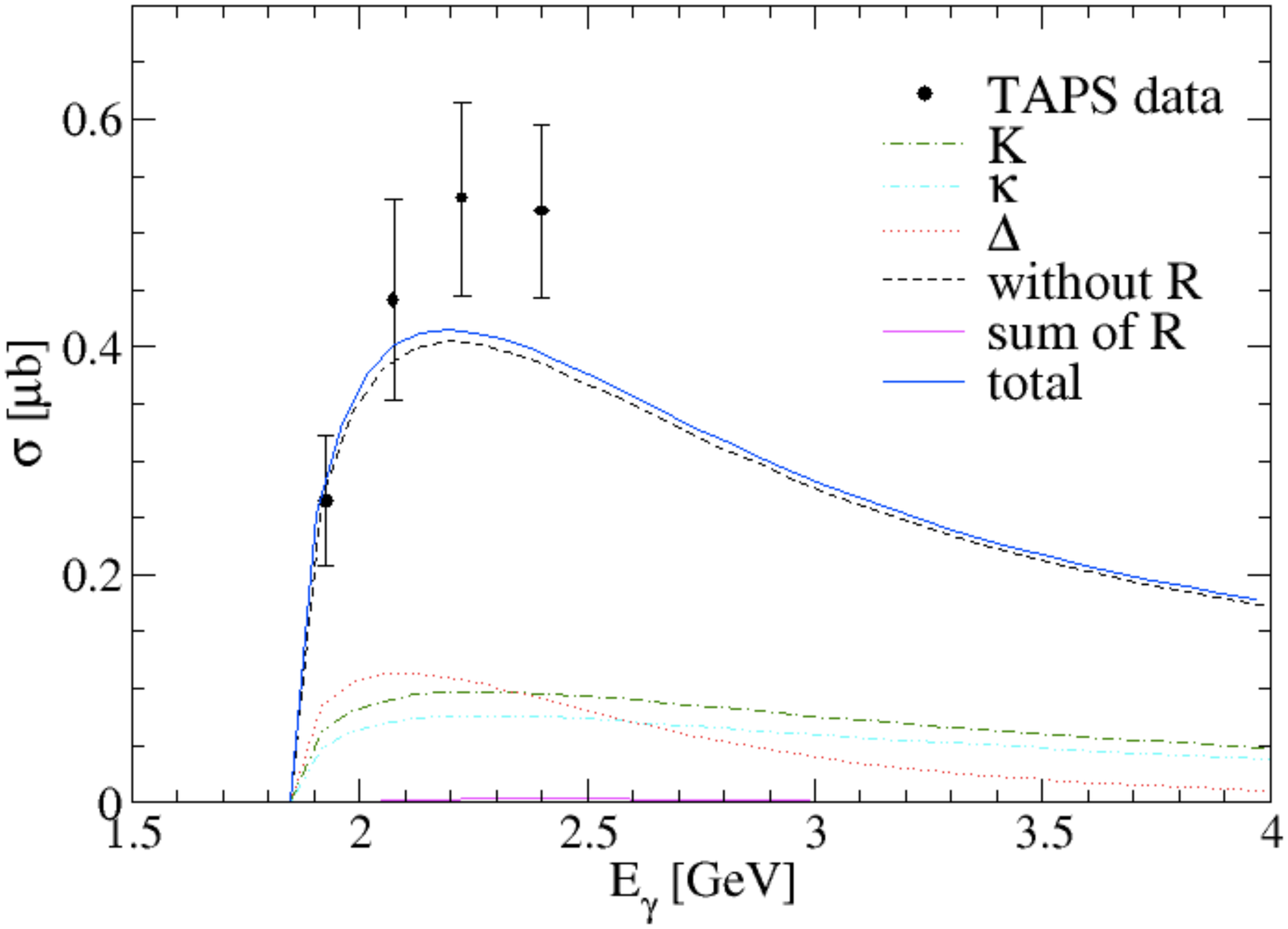}
\includegraphics[width=8.5cm]{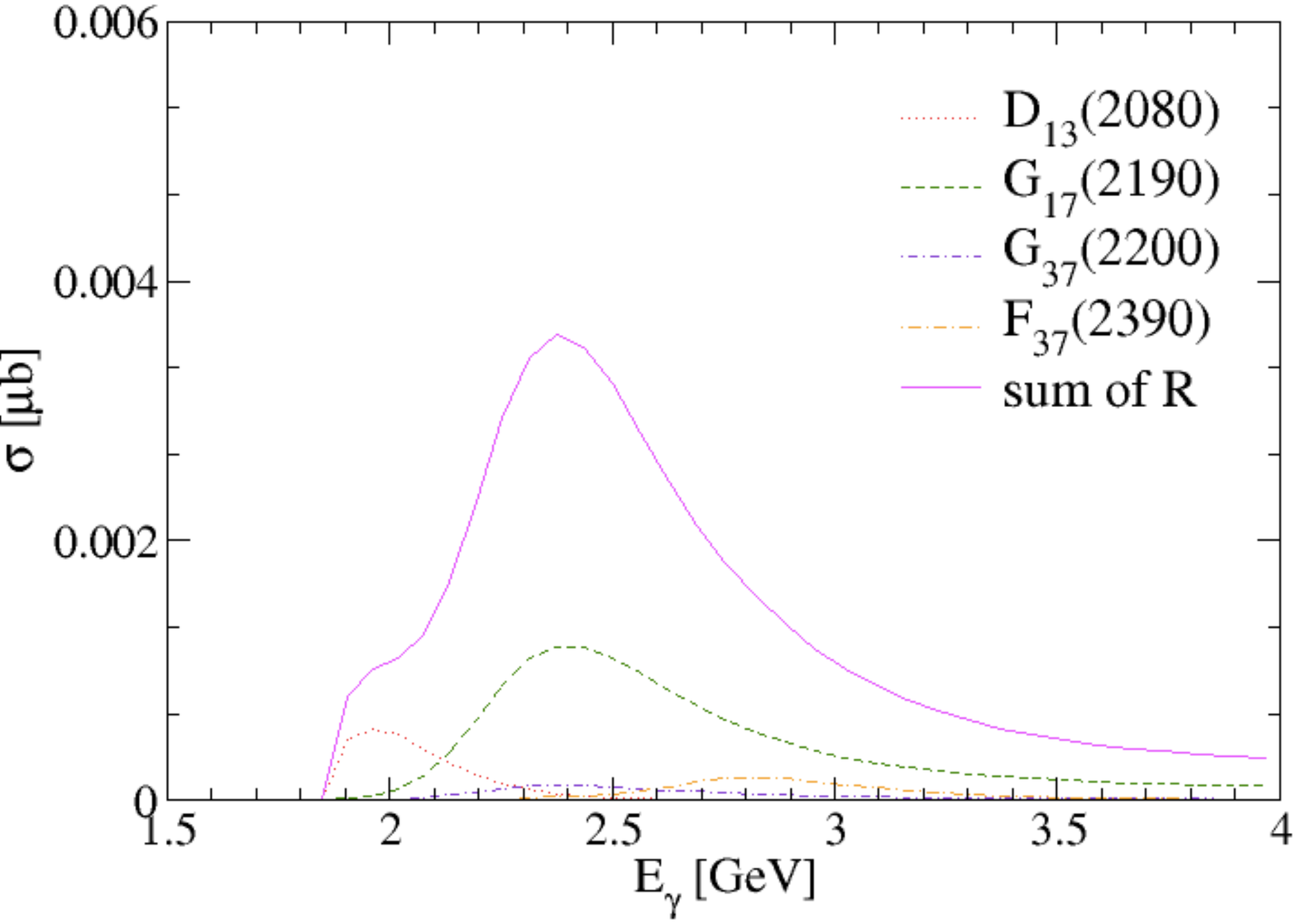}
\end{tabular}
\caption{(Color online) Left: Total cross sections as functions of 
  $E_\gamma$. The experimental data are taken from the
  TAPS~\cite{Nanova:2008kr}. Right: Total cross sections from the
  resonance contributions in the $s$ channel.}        
\label{FIG3}
\end{figure}

The numerical results for the total cross sections are given in the
left panel of Fig.~\ref{FIG3} as functions of $E_\gamma$. Note that
the numerical results are slightly underestimated for
$E_\gamma=(2\sim2.5)$ GeV in comparison to the data even with the
resonance contributions.  In the right panel of Fig.~\ref{FIG3}, we
depict each contribution from the $s$-channel resonances. Although
$D_{13}$ and $G_{17}$ provide sizable contributions near the
threshold, they are still far smaller than those from the Born
contributions. We note that this tendency of the small effects from
the resonances are obviously different from the
$K^*\Lambda$~\cite{Oh:2006hm,Kim:2011rm} and
$K\Lambda$~\cite{Janssen:2001pe} photoproductions. The difference
between the $\Lambda$ and $\Sigma$ channels can be understood by the
much smaller strong couplings for the $\Sigma$ channel computed by the
SU(6) quark model in comparison to those for the $\Lambda$
channel~\cite{Capstick:1998uh}.  

\section{Summary}
We have investigated the $K^{*0}\Sigma^+(1190)$ photoproduction using
the effective Lagrangian method at the tree-level Born
approximation. In addition to the Born terms, we also took into
account the resonance contributions in the $s$ and $u$ channels. The
experimental data were reproduced qualitatively well. It also turns
out that the resonance contributions are almost negligible in the
energy and angular dependences of the cross sections, being different
from the $K\Lambda$ and $K^*\Lambda$ photoproductions. The
strange meson ($K$ and $\kappa$) exchanges in the $t$ channel and
the $\Delta(1232)$ contribution in the $s$ channel are the most
dominant in describing the $K^{*0}\Sigma^+(1190)$
photoproduction. More detailed works are under progress and appear
elsewhere.   

\section*{Acknowledgments}
This talk was presented at the 20th International IUPAP Conference on Few-Body Problems in Physics, $20 \sim25$ August 2012, Fukuoka, Japan. The authors thank Y.~Oh and K.~Hicks for fruitful discussions on this subject. The present work is supported by Basic Science  Research Program through the National Research Foundation of Korea funded by the Ministry of Education, Science and Technology (Grant Number: 2012R1A1A2001083).

\end{document}